\documentclass[prl,aps,showpacs,twocolumn,floatfix]{revtex4}
\usepackage{epsfig} \usepackage{graphics} \usepackage{bm}
\usepackage{amssymb}
\begin{document}

\title{Is Gravitational Mass of a Composite Quantum Body
Equivalent to its Energy?}

\author{A.G. Lebed$^*$}

\affiliation{Department of Physics, University of Arizona, 1118 E.
4-th Street, Tucson, AZ 85721, USA (lebed@physics.arizona.edu)}
\date{\today}

\begin{abstract}
We define gravitational mass operator of a hydrogen atom in the post-Newtonian approximation of the General Relativity and show that it does not commute with 
energy operator. 
Nevertheless, the equivalence between the expectation values of gravitational mass 
and energy is shown to survive for stationary quantum 
states.
Inequivalence between gravitational mass and energy at a microscopic level reveals 
itself as unusual electromagnetic radiation, emitted by the atoms, supported and 
moved in the Earth gravitational field, which can be experimentally 
measured. 
Inequivalence between gravitational mass and energy at a macroscopic level results 
in time dependent oscillations of the expectation values of gravitational mass for 
mixed quantum states.
\end{abstract}

\maketitle

\maketitle

One of the major problems in physics is known to be a creation of
the so-called "Theory of Everything" - unification of all
fundamental forces in nature, including electro-weak, strong, and
gravitational ones. In this context, the most difficult step is a
creation of quantum gravitational theory. 
This even may not be possible in the feasible future, since the fundamentals 
of the quantum mechanics are very different from that of the General Relativity 
(GR).
In this situation, it is important to find a way to combine the quantum
mechanics with some non-trivial approximation of the 
GR. 
This allows to introduce some novel physical ideas and phenomena, which
can be hopefully experimentally studied. 
So far, to the best of our knowledge, only trivial quantum mechanical variant 
of the Newton approximation of the GR has been experimentally tested in the 
famous COW \cite{COW-1} and ILL \cite{ILL-1} 
experiments. 
On the other hand,
such important and non-trivial quantum phenomena in the GR as the
Hawking radiation \cite{Hawking} and the Unruh effect \cite{Unruh}
are still far from their direct and firm experimental confirmations.

It is known that gravitational mass of a composite classical body in the GR is
not a trivial notion and is a subject of several 
paradoxes. 
One of them is related to application of the so-called Tolman formula
\cite{Tolman} to a free photon, which formally results in a doubling
of photon active gravitational mass \cite{Tolman, Misner}. 
The solution of this paradox is due to an account of stress in the
walls of a container \cite{Misner}, containing photon, which
compensates the above mentioned increase of 
photon mass.
More precisely, it is shown \cite{Misner} that averaged over time
gravitational mass of a photon in a container with mirror walls is
$E/c^2$ , where $E$ is photon 
energy.
Importance of the classical virial theorem for the equivalence
between averaged over time gravitational mass and energy
of different composite classical bodies is stressed in
Refs.\cite{Nordtvedt,Carlip}. 
In particular, it is shown that electrostatic binding energy, $U$, contributes 
to passive and active gravitational masses as $2U/c^2$, whereas kinetic energy,
$K$, contribution to the gravitational masses is $3K/c^2$
\cite{Nordtvedt,Carlip}. 
Application of the classical virial theorem, which claims that  averaged over
time potential energy, $<U>_t$, equals to -$2<K>_t$ in the above 
mentioned case, results in the equivalence between averaged over time 
gravitational mass and
energy, $<m>_t=E/c^2$.

The main goal of our Letter is to consider a quantum mechanical
problem about interaction of a composite body (e.g., a hydrogen
atom) with an external gravitational field 
(e.g., the Earth). 
The first our result is that the equivalence between gravitational mass
and energy may survive at a macroscopic level. 
In particular, we show
that the quantum virial theorem \cite{Virial} results in equality
between the expectation value of gravitational mass and energy,
divided into $c^2$, for stationary quantum 
states. 
The second our result is a breakdown of the equivalence between 
gravitational mass and energy at a microscopic
level. 
In the Letter, we define gravitational mass operator of a hydrogen
atom in the post-Newtonian approximation of the GR and show that it
does not commute with energy operator. 
Therefore, an atom with definite energy $E$ is not characterized by definite 
gravitational mass in an external gravitational 
field. 
Gravitational mass is shown to be quantized and can significantly differ 
from $E/c^2$. 
We discuss how the above mentioned phenomenon can be
experimentally observed. 
In particular, we suggest experimental investigation of electromagnetic 
radiation, emitted by macroscopic ensemble of the atoms, supported and 
moved with constant velocity in the Earth
gravitational field \cite{Lebed-1}.
The third our result is that the equivalence between the expectation values 
of gravitational mass and energy is broken at macroscopic level for mixed 
quantum states.

Here, we derive some approximation of the Dirac equation for a
hydrogen atom in the Earth gravitational field.
 It is well known that the interval in a weak central symmetric gravitational field
can be written as \cite{Misner-2}
\begin{eqnarray}
d s^2 = -\biggl(1 + 2 \frac{\phi}{c^2} \biggl)(cdt)^2 
+ \biggl(1 - 2 \frac{\phi}{c^2} \biggl) (dx^2 +dy^2+dz^2 ),
\nonumber\\
\phi = - \frac{GM}{R} ,
\end{eqnarray}
where $G$ is the gravitational constant, $c$ is the velocity of
light, $M$ and $R$ are the Earth mass and radius, respectively. 
Let us introduce the proper local coordinates and time,
\begin{eqnarray}
&&x'=\biggl(1-\frac{\phi}{c^2} \biggl) x, \ y'= \biggl(1-\frac{\phi}{c^2} \biggl) y, 
\nonumber\\
&&z'=\biggl(1-\frac{\phi}{c^2} \biggl) z , \ t'= \biggl(1+\frac{\phi}{c^2} \biggl) t,
\end{eqnarray}
where the interval has the Minkowski form,
\begin{equation}
(d s')^2 = -(cdt')^2 + (dx')^2 + (dy')^2 +(dz')^2 .
\end{equation}
In this case, we can approximately write the Dirac equation in the proper 
coordinates and time as
\begin{equation}
i \hbar \frac{\partial \Psi'}{\partial t'} = \hat H' \Psi'  , \  \
\hat H' =  c \bm{\alpha} \hat {\bf p'}  + \beta m_e c^2  - \frac{e^2}{r'},
\end{equation}
where $\bm{\alpha}$ and $\beta$ are the standard matrices, vector
${\bf r'}$ is characterized by coordinates $(x',y',z')$, $\hat {\bf
p'} = -i \hbar \partial / \partial {\bf r'}$; $m_e$ is the bare
electron mass, e is the electron charge. 

The next two steps are to take the non-relativistic Pauli approximation of 
Eq.(4) and to rewrite it in terms of the coordinates $(x, y, z)$ and 
time $t$. 
As a result, we obtain the following non-relativistic
Hamiltonian, which contains couplings of the Earth gravitational
field with kinetic and potential energies of an electron:
\begin{equation}
\hat H = \frac{\hat {\bf p}^2}{2m_e}-\frac{e^2}{r} + \hat
m^g_e \phi \ ,
\end{equation}
where we introduce electron gravitational mass operator,
\begin{equation}
\hat m^g_e =m_e+ \biggl(\frac{\hat {\bf
p}^2}{2m_e}-\frac{e^2}{r}\biggl)/c^2 + \biggl(2 \frac{\hat {\bf
p}^2}{2m_e}-\frac{e^2}{r} \biggl)/ c^2 \ .
\end{equation}
Note that the first term in Eq.(6) is the bare electron mass, $m_e$, 
the second term corresponds to the expected electron energy
contribution to gravitational mass operator, whereas the third
non-trivial term is the virial contribution to gravitational
mass operator.
It is important that a comparison of Eqs.(5),(6) with Refs. 
\cite{Fischbach, Obukhov} shows that the suggested above approximated 
method of a derivation of the Dirac equation in a curved spacetime 
disregards only tidal effects \cite{Lebed-2}.

Suppose that we have a macroscopic ensemble of hydrogen atoms with each 
of them being in a ground state with energy $E_1$. 
Then, it follows from Eq.(6) that the expectation value of electron gravitational 
mass is
\begin{equation}
<\hat m^g_e> = m_e+ E_1/c^2 + \biggl< 2 \frac{\hat
{\bf p}^2}{2m_e}-\frac{e^2}{r} \biggl>/ c^2 = m_e + E_1/c^2 \ ,
\end{equation}
where the third term in Eq.(7) gives zero in accordance with
the quantum virial theorem \cite{Virial}. 
Therefore, we conclude that
the equivalence between gravitational mass and energy survives
at a macroscopic level for stationary quantum states.
In other words, in this case the expectation value of gravitational mass 
is equal to the total energy, divided into $c^2$.
Let us discuss how Eqs.(5),(6) break the equivalence between gravitational
mass and energy at a microscopic level.
First of all, we pay attention that the gravitational mass operator (6) does
not commute with electron energy operator in the absence of gravitational
field.
This means that, if we create a quantum state of a hydrogen atom with
definite energy, it will not be characterized by definite gravitational
mass.
In other words, a measurement of gravitational mass in such quantum
state may give different values, which, as shown, are quantized.

Here, we illustrate the breakdown of the equivalence between gravitational 
mass and energy at a microscopic level, using the following thought 
experiment.
Suppose that we create a quantum state of a hydrogen atom with definite energy
in the absence of a gravitational field and then adiabatically switch on the
gravitational field (1).
More specifically, at $t \rightarrow - \infty$ (i.e., in the absence of gravitational field),
a hydrogen atom is in its ground state with wave 
function,
\begin{equation}
\Psi_1(r,t) = \Psi_1(r) \exp(-iE_1t/\hbar) \ ,
\end{equation}
whereas in the vicinity of $t=0$ [i.e., in the presence of the gravitational field(1)] it is characterized by the following
wave function:
\begin{equation}
\Psi(r,t) = \sum^{\infty}_{n = 1} a_n (t) \Psi_n(r) \exp(-i E_n t/\hbar) \ .
\end{equation}
[Here, $\Psi_n(r)$ is a normalized wave function of an electron in a hydrogen
atom in the absence of a gravitational field, corresponding
to energy $E_n$ \cite{Lebed-3}.]

As it follows from Eqs.(5),(6), adiabatically switched on gravitational field corresponds
to the following time-dependent small
perturbation:
\begin{equation}
\hat U ({\bf r},t) = \phi \exp(\lambda t) \biggl[ \biggl(\frac{\hat {\bf p}^2}{2m_e}-\frac{e^2}{r}\biggl)/c^2
+ \biggl(2 \frac{\hat {\bf p}^2}{2m_e}-\frac{e^2}{r} \biggl)/ c^2 \biggl] ,
\end{equation}
where $\lambda \rightarrow 0$.
The standard calculations by means of the time-dependent quantum mechanical perturbation
theory give the following results:

\begin{equation}
a_1(t)=\exp \Big( - \frac{i \phi E_1  t}{c^2 \hbar} \Big) \  ,
\end{equation}

\begin{equation}
a_n(0)=  - \frac{\phi}{c^2} \frac{V_{n,1}}{E_n-E_1} \  , \ n \neq 1 \ ,
\end{equation}
where
\begin{equation}
V_{n,1}= \int \Psi^*_n(r) \hat V({\bf r}) \Psi_1(r) d^3 {\bf r} \ ,
\end{equation}
with the virial term being
\begin{equation}
\hat V({\bf r}) = 2 \frac{{\bf \hat p}^2}{2 m_e} - \frac{e^2}{r}\ .
\end{equation}
[Note that the perturbation (10) is characterized by the following selection 
rule.
Electron from $1S$ ground state of a hydrogen atom can be excited
only into $nS$ excited state.] 

Let us discuss Eqs.(11)-(14).
It is important that Eq.(11) corresponds to the well-known red shift of atomic
ground state energy $E_1$ in the gravitational field (1).
On the other hand, Eq.(12) demonstrates that there is a finite probability,
\begin{equation}
P_n = |a_n(0)|^2 = \Big( \frac{\phi}{c^2} \Big)^2 \
\Big(  \frac{V_{n,1}}{E_n-E_1} \Big)^2 \ , \ n \neq 1,
\end{equation}
that at $t=0$ electron occupies n-th energy 
level.
In fact, this means that measurement of gravitational mass (6)
in a quantum state with definite energy (8) gives the following
quantized values:
\begin{equation}
m^g_e (n) = m_e + (E_n-E_1)/c^2 \ ,
\end{equation}
with the probabilities (15) for $n \neq 1$.
Note that, although the probabilities (15) are quadratic with respect to
gravitational potential and, thus, small,  the corresponding changes of 
gravitational mass (16) are large and of the order of $\alpha^2 m_e$, 
where $\alpha$ is the fine
structure constant.
It is important that the excited levels of a hydrogen atom spontaneously decay,
therefore, one can detect the above discussed quantization law of gravitational
mass (16) by measuring electromagnetic radiation, emitted by a macroscopic
ensemble of hydrogen atoms.

Below, we discuss a more realistic experiment.
Suppose that a hydrogen atom is in its ground state (8),
\begin{equation}
\tilde{\Psi}_1(r, t) = \Psi_1[(1-\phi'/c^2)r] \exp[-iE_1(1+\phi'/c^2)t/\hbar] \ ,
\end{equation}
at $t=0$ and located in area, where the Earth gravitational potential is
$\phi' = \phi(R')$, where $R'$ is a distance between a hydrogen atom and center
of the Earth.
Suppose that a hydrogen atom is supported and moved from the Earth with
constants velocity,  $v \ll \alpha c$, where $\alpha c$ is a characteristic electron 
velocity in a hydrogen 
atom.
In this case, as it follows from Ref.[7], the extra so-called "gravimagnetic" 
contributions to the Lagrangian [and, thus, to the Hamiltonian (5),(6)] are 
small in an inertial system, related to a hydrogen
atom.
Therefore, time dependent perturbation  for the Hamiltonian in the above 
mentioned inertial coordinate system can be written as
\begin{equation}
\hat U ({\bf r},t) =\frac{\phi(R'+vt)-\phi(R')}{c^2}
 \biggl(3 \frac{\hat {\bf p}^2}{2m_e}-2\frac{e^2}{r} \biggl) .
\end{equation}
In this case, the time-dependent quantum mechanical perturbation theory
gives the following solution for electron wave functions:

\begin{equation}
\tilde{\Psi}(r,t) = \sum^{\infty}_{n=1} \tilde{a}_n(t) \Psi_n[(1-\phi'/c^2)r] \exp[-iE_n(1+\phi'/c^2)t/\hbar] ,
\end{equation}
\begin{equation}
\tilde{a}_1(t)=\exp \Big\{ - i \frac{E_1}{\hbar c^2} \int^t_0
[\phi(R'+ut) - \phi(R')] dt \Big\}  \  ,
\end{equation}
\begin{equation}
\tilde{a}_n(t)= \frac{\phi(R') - \phi(R'+vt)}{c^2} \frac{V_{n,1}}{E_n-E_1}
\exp(i \omega_{n,1} t) \  , \ n \neq 1 \ ,
\end{equation}
where $\omega_{n,1}=(E_n - E_1) / \hbar$.

It is important that, if the excited levels of a hydrogen atom were strictly stationary, 
then a probability to find gravitational mass to be quantized with $n \neq 1$ in 
Eq. (16) would be
\begin{equation}
\tilde{P}_n(t)=  \frac{[\phi(R') - \phi(R'+vt)]^2}{c^4} 
\biggl( \frac{V_{n,1}}{E_n-E_1} \biggl)^2.
\end{equation}
In reality, the excited levels spontaneously decay, therefore, it is possible
to observe the quantization law for gravitational mass (16) indirectly by
measuring electromagnetic radiation from a macroscopic ensemble of
the atoms.
In this case, Eq.(22) gives a probability that a hydrogen atom emits a photon
with frequency $\omega_{n,1} = (E_n-E_1) / \hbar$ during time interval
$t$ \cite{Lebed-4}.

Let us estimate the probability (22).
If the experiment is done by using satellite, then we may have
$|\phi(R'+ut)| \ll |\phi(R')|$.
Under such condition equation (22) coincides with Eq. (15):
 \begin{equation}
\tilde{P}_n = \frac{\phi^2(R')}{c^4} \biggl( \frac{V_{n,1}}{E_n-E_1} \biggl)^2
 \simeq  0.49 \times 10^{-18} \biggl( \frac{V_{n,1}}{E_n - E_1} \biggl)^2 ,
\end{equation}
where we use the following  numerical values of the Earth mass,
$M \simeq 6 \times 10^{24} \ kg$, and its radius,  $R \simeq 6.36 
\times 10^6 \ m$.
Note that, although the probability (23) is small, the number of photons, 
N, emitted by macroscopic ensemble of the atoms, can be large since the factor
$V^2_{n,1}/ (E_n - E_1)^2$ is of the order 
of unity.
For instance, for 1000 moles of hydrogen atoms the number of emitted photons 
is estimated as
\begin{equation}
N(n \rightarrow 1) = 2.95 \times 10^{8} \biggl( \frac{V_{n,1}}{E_n-E_1} \biggl)^2 ,
\end{equation}
which can be hopefully experimentally detected 
\cite{Lebed-1,Lebed-4}.
In particular, the number of emitted photons for the quantum transition 
$2S \rightarrow 1S$ 
is equal to
\begin{equation}
N(2 \rightarrow 1) = 0.9 \times 10^{8} .
\end{equation}
Note that the experiments on a detection of photons (22)-(25) have to be done at 
low enough temperatures, where the number of temperature activated electrons is negligible.
This seems not to be a very difficult problem due to exponential dependence of
their number on temperature.

To summarize, we have shown that gravitational mass of a composite 
quantum body is not equivalent to its energy due to quantum fluctuations 
and suggested experimental method to detect the corresponding 
difference.
This clear demonstrates inequivalence of gravitational mass and energy at a 
microscopic level.
On the other hand, we have shown that the expectation value of gravitational 
mass of an ensemble of hydrogen atoms in their ground states is equal 
to $E_1/c^2$ per atom.
In this context, we would like to make the following comments.
First of all, we stress that, for mixed states, the expectation value of 
gravitational mass can be oscillatory function of time even in case, where the 
expectation value of energy is 
constant. 
Indeed, as it follows from Eq.(6), for electron wave function,
\begin{equation}
\Psi_{1,2}(r,t) = \frac{1}{\sqrt{2}} \bigl[ \Psi_1(r) \exp(-iE_1t) + \Psi_2(r) \exp(-iE_2t) \bigl],
\end{equation}
which is characterized by the time independent expectation value of energy,
$<E> = (E_1+E_2)/2$,
the expectation value of gravitational mass is the following oscillatory function 
\cite{Lebed-5}:
\begin{equation}
<\hat m^g_e> = m_e + \frac{E_1+E_2}{2 c^2} + \frac{V_{1,2}}{c^2}
\cos \biggl[ \frac{(E_1-E_2)t}{\hbar} \biggl] .
\end{equation}

Note that Eq.(27) directly demonstrates inequivalence between gravitational
mass and energy at a macroscopic level.
In the latter case, only the averaged over time expectation value of gravitational 
mass of a composite quantum body is equivalent to its 
energy.
It is important that gravitational mass oscillations (27) is of a pure quantum mechanical nature and do not have their classical 
analog.
In this context, we pay attention that the averaging procedure of oscillations (27)
has completely different physical meaning than the averaging procedure in a
classical case, discussed in the beginning of the Letter.  
We hope that oscillations (27) can be experimentally discovered in the future, 
despite the fact that the atomic state (26) spontaneously decays with time due 
to emission of photons.

In conclusion, we pay attention that all suggested in the Letter phenomena are
very general and due to non-zero curvature of a space in the presence of
a gravitational field.
In particular, electromagnetic energy scale disappears from Eqs.(22)-(25).
Therefore, we expect that the analogous effects may exist if we consider a nucleus
or an elementary particle in an external gravitational
field.

We are thankful to N.N. Bagmet for useful discussions.
This work was supported by the NSF under Grants DMR-0705986 and
DMR-1104512.

$^*$ Also, at the L.D. Landau Institute for Theoretical Physics, 2
Kosygina Street, 117334 Moscow, Russia


\begin{references}
\bibitem{Lebed-1}
In practice, it may be more convenient to investigate electromagnetic
radiation, emitted by some solid body.
\bibitem{Lebed-2}
Note that we keep in the Hamiltonian (5),(6) only terms of the order of 
$1/c^2$ and, thus, take into account only couplings of kinetic and electrostatic
potential energies of an electron with an external gravitational field.
\bibitem{Lebed-3}
Due to symmetry of our problem, we need to keep in Eq.(9) only wave
functions, corresponding to $nS$ quantum states.
\bibitem{Lebed-4}
We pay attention that $nS \rightarrow 1S$ quantum transitions can exist only 
due to quadrupole effects, since dipole matrix elements for such transitions
are zero.
\end{references}
\end{document}